\documentclass[useAMS,usenatbib]{mn2e}
\usepackage{graphicx}
\usepackage{amsmath}
\title[New SX Phe variables in the globular cluster NGC\,288]{New SX Phe variables in the globular cluster NGC\,288\footnote{Based on observations obtained at the Southern Astrophysical Research (SOAR) telescope, which is a joint project of the Minist\'erio da Ci\^encia, Tecnologia, e Inova\c{c}\~ao (MCTI) da Rep\'ublica Federativa do Brasil, the U.S. National Optical Astronomy Observatory (NOAO), the University of North Carolina at Chapel Hill (UNC), and Michigan State University (MSU).} }

\author[E.~Martinazzi, S. O. Kepler, J. E. S.~Costa, A.~Pieres, C.~Bonatto, E.~Bica and L.~Fraga]{E.~Martinazzi$^{1,2}$\thanks{E-mail: elizandra.martinazzi@ufrgs.br}, 
S.~O.~Kepler$^{1}$, J.~E.~S.~Costa$^{1}$, A.~Pieres$^{1}$, C.~Bonatto$^{1}$, \newauthor E.~Bica$^{1}$ and L.~Fraga$^{3}$ \\
$^{1}$Instituto de F\'{\i}sica, Universidade Federal do Rio Grande do Sul, 91501-900 Porto Alegre, RS, Brazil\\
$^{2}$Instituto Federal do Rio Grande do Sul, 95700-000, Bento Gon\c{c}alves, RS, Brazil \\
$^{3}$Laborat\'orio Nacional de Astrof\'{\i}sica – LNA/MCTI, R. Estados Unidos, 154, Itajub\'a, 37504-364, MG, Brazil}

\begin{document}

\date{Accepted, 16 March 2015. Received, 15 March 2015; in original form, 10 February 2015}

\pagerange{\pageref{firstpage}--\pageref{lastpage}} \pubyear{XXXX}

\maketitle

\label{firstpage}

\begin{abstract}
We report the discovery of two new variable stars in the metal-poor globular cluster NGC\,288, 
found by means of time-series CCD photometry. We classified the new variables as SX Phoenicis due to their characteristic fundamental mode periods ($1.02 \pm 0.01$ and $0.69 \pm 0.01$ hours), and refine the period estimates for other six known variables. SX Phe stars are known to follow a well-defined Period-Luminosity (P-L) relation and, thus, can be used for determining distances; they are more numerous than RR Lyraes in NGC\,288. We obtain the P-L relation for the fundamental mode 
M$_V$ = ($-2.59 \pm 0.18$) $\log P_0$(d) + ($-0.34 \pm 0.24$) and for the first-overtone mode 
M$_V$ = ($-2.59 \pm 0.18$) $\log P_1$(d) + ($0.50 \pm 0.25$). 
Multi-chromatic isochrone fits to our UBV color-magnitude diagrams, based on the Dartmouth Stellar Evolution Database, provide $\langle$[Fe/H]$\rangle$ = -1.3 $\pm$ 0.1, E(B-V) = 0.02 $\pm$ 0.01 and absolute distance modulus (m-M)$_{0}$ = 14.72 $\pm$ 0.01 for NGC\,288.
\end{abstract}


\section{Introduction}

The study of stellar systems as the globular clusters provides important clues to the Galaxy formation history. Each cluster is made up by a relatively simple population of stars, i.e. practically all stars were born coeval, in the same region, and from the same molecular cloud \citep[e.g.][]{Rosenberg2000}. 
In latter decades several studies prove the presence of two (or even more) populations of stars, setting a puzzling genesis for these objects \citep{2012ApJ...760...39P}. Globular clusters are among the oldest objects in the Galaxy, and their ages provide basic information on the early stages of Galactic formation \citep[e.g.][]{Gratton2003}.

The Galactic globular cluster NGC\,288 is placed at RA= 00$^{h}$52$^{m}$45$^{s}$.24 and Dec.= -26$^{o}$34$\arcmin$57$\arcsec$.4 (J2000), close to the South Galactic Pole, with Galactic coordinates (J2000): \textit{l} = 152.$^{o}$28 and \textit{b} = -89.$^{o}$38 \citep{2010AJ....140.1830G}. 
It is located approximately 8.9\,kpc from the Sun, in a region of small interstellar extinction, with quite low reddening [E(B-V) = 0.03] and a published distance modulus of (m-M)$_{V}$ = 14.84 [\cite{Harris1996} (2010 edition)].  NGC\,288 is a globular cluster with a fairly low central density and is relatively metal-poor. 

NGC\,288 has a well-marked blue horizontal branch \citep{2014A&A...565A.100M} and presents the multiple-population phenomenon, exhibiting two distinct red giant branches (RGBs), two populations of stars characterized by difference in  light-element abundance \citep{2009PASP..121.1054S}.
The possible metallicity spread is evidenced by a split red giant branch (RGB), observed in  color-magnitude diagrams (CMDs), that may be best-fitting by isochrone stars whose second generation is $\sim$ 1.5\,Gyr younger \citep{2014arXiv1406.5220H,2011ApJ...733L..45R}. 
\cite{2013ApJ...775...15P} used multi-band HST photometry covering a wide range of wavelengths and found that NGC\,288 main sequence splits into two branches and that this duality is repeated along the subgiant branch (SGB) and the RGB, consistent with two distinct stellar populations.

We report a search for variable stars in NGC\,288.

\section{Observations and data reduction}

We obtained the NGC\,288 CCD photometric data with the 4.1-m SOAR telescope in 2\,013 October 28, using the Imaging Goodman Spectrograph \citep{2004SPIE.5492..331C}. 
The field was centred at RA= 00$^{h}$52$^{m}$44$^{s}$.9 and Dec.= -26$^{o}$34$\arcmin$57$\arcsec$.4 (J2000), approximately on the cluster centre. In imaging mode the field of view is 7.2\,arcmin in diameter, with 1\,548$\times$1\,548 pixels and plate scale of 0.30\,arcsec$\cdot$pixel$^{-1}$ when binned by 2$\times$2.

We obtained 300 useful images in the B-Bessel filter, using exposure times of 60\,s, with a total of $\sim$7\,hours of observation. 
To construct the CMD, we obtained images with exposure times of 1, 10, 120 and 300\,s for the V and B filters, and 300 and 600\,s for the U filter.

Data reduction were carried out with \textsc{IRAF}.  We used tasks of the \textsc{IRAF} \textit{daophot} package for crowded-field stellar photometry \citep{1987PASP...99..191S}: \textsc{daofind} to find stars in each image, \textsc{phot} to compute sky values and initial magnitudes for the found stars and \textsc{allstar} to group and fit point spread function (PSF) to all stars simultaneously.  The PSF best-fitting was obtained with a elliptical Moffat function with $\beta = 2.5$. 

In total, were generated light curves for 11\,389 sources inside the field and with B magnitude down to 24.  Differential photometry was performed, computed by taking the difference between the instrumental magnitudes of an unknown star and of a known standard star with constant brightness \citep[e.g.][]{1998IAPPP..72...42B} inside the same field, to determine how the stellar brightness changes in time. 

\section{NGC\,288 parameters}

\begin{figure*}
\begin{minipage}{115mm}
\centering
\includegraphics[width=115mm,clip=]{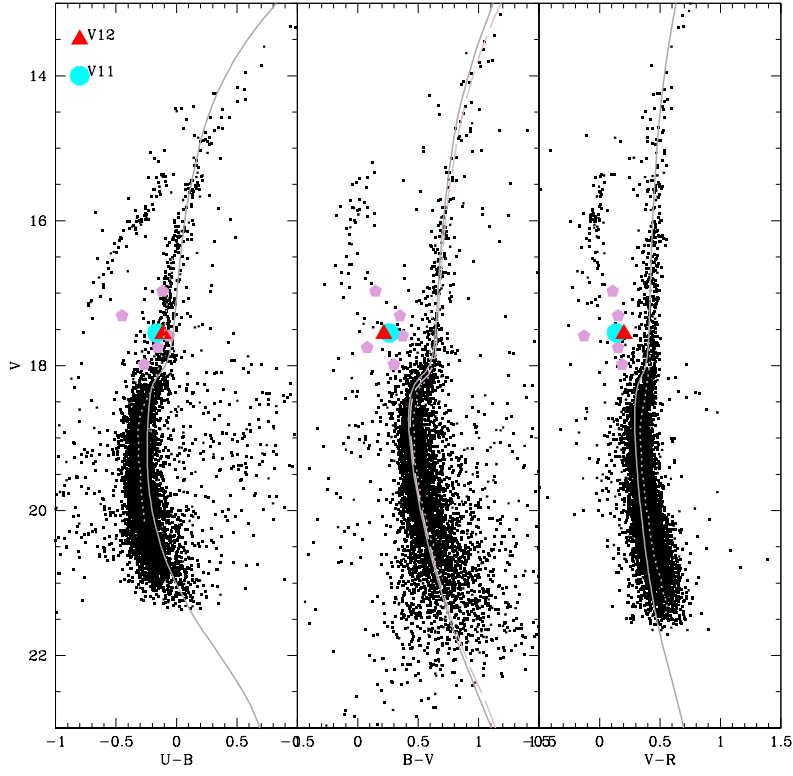}
\caption{ Color-magnitude diagrams for NGC\,288 from 4.1-m SOAR telescope with U, B and V filters, showing the location of variable stars. The dotted line is the mean ridge line. 
The continuous line is the best fitting multi-chromatic isochrone. The parameters for NGC\,288 [Fe/H] = -1.3 $\pm$ 0.1, E(B-V) = 0.02 $\pm$ 0.01 
magnitudes, distance modulus (m-M)$_{0}$ = 14.72 $\pm$ 0.01 magnitudes and label of age 13.5\,Gyr. The CMD in B-V was corrected by the reddening map, shown in Fig.~\ref{N288_HR_DRmap}. The dashed line shows the best fit for the B-V color 
only, with [Fe/H] = -1.2 $\pm$ 0.1 and distance modulus (m-M)$_{0}$ = 14.72 $\pm$ 0.01. }
\label{CMDMart}
\end{minipage}
\end{figure*}

\begin{figure*}
\begin{minipage}{115mm}
\centering
\includegraphics[width=115mm,clip=]{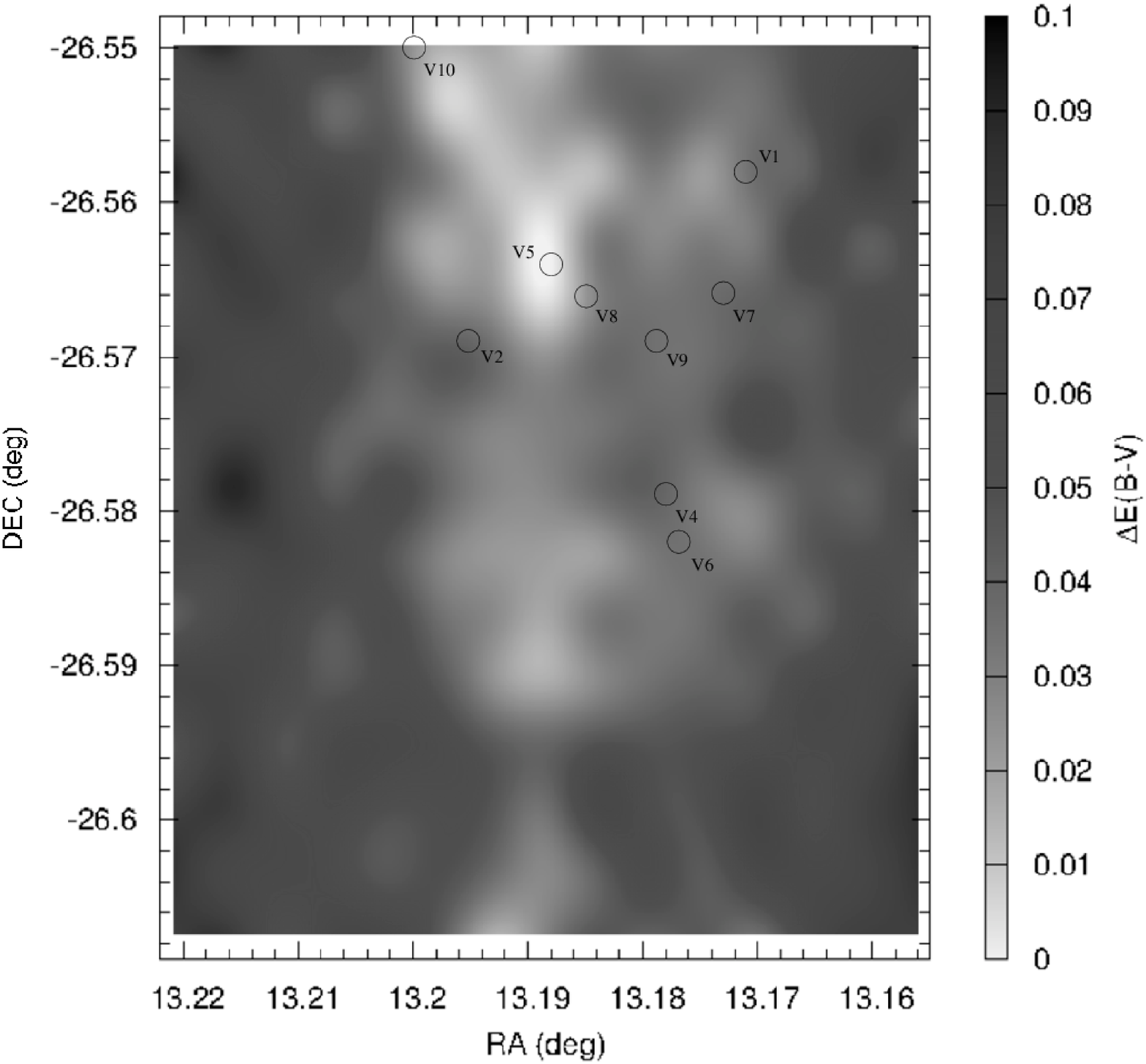}
\caption{ Differential Reddening (DR) map: difference E(B-V) in cell RA$\times$DEC (degrees) with respect to E(B-V) of the cell with lower extinction. Since NGC\,288 has low frontal extinction, the DRs represent almost exactly the value for E(B-V). }
\label{N288_HR_DRmap}
\end{minipage}
\end{figure*}

The CMDs for NGC\,288 from the 4.1-m SOAR telescope with U, B, V and R filters are shown in Fig.~\ref{CMDMart}. 
For the CMD with V magnitude and (B-V) color, we made a correction using the differential-reddening (DR) map in Fig.~\ref{N288_HR_DRmap}, built according to \cite{2013MNRAS.435..263B}. This map is based on colour shifts among stars extracted from different regions across the field of NGC\,288. As a caveat, we note that stars extracted from wide-apart regions may present a difference in colour related to zero-point variations across the field, which would be taken as a difference in reddening. Zero-point variation across fields may be related to inaccuracies on the PSF model, sky/bias determination, or sky concentration \citep[e.g.][]{2012A&A...540A..16M}. So, part of the differential reddening in Fig.~\ref{N288_HR_DRmap} may in fact be residual zero-point variations.

We used a statistical approach to determine fundamental parameters of the globular cluster NGC\,288 (age, metallicity, reddening and distance modulus). 
To proceed with this method we determine a Mean Ridge Line (MRL) to the set of stars in each CMD. This fiducial line is determined as a group of points where each point represents an average of stars in CMD plane. The distance between each couple of points in fiducian line is around 0.1 magnitude (in the CMD plane) and is an averaged value from ten other local mean points, using small shifts to avoid local deviations (lack or over density of stars in CMD plane). To avoid deviations from field stars and binaries, we perform a sigma-clip five times. The uncertainties from this method was determined using simulations and reaches around 0.02 magitude even with a 100 per cent binaries and a 600 stars population. To fit NGC\,288 parameters, we compared these fiducial lines in color-magnitude diagrams (V against B-V and V-I colours) with grids of Dartmouth Stellar Evolution Database \citep{2008ApJS..178...89D}. 

Analyzing the individual fits, the metallicity sensitivity is greatly impaired for redder colors. 
On the other hand, age nearly degenerates for bluer filters. The mean metallicity estimated from the CMD using this method agrees with the spectroscopic metallicity, [Fe/H] = -1.1 $\pm$ 0.1 (without the U-B color). The model age of least dispersion is 13.5\,Gyr. The reddening is [E(B-V)]=0.05 and the distance modulus of best agreement is 14.57 $\pm$ 0.08  for the V filter. 
This results agree with reddening  [E(B-V)=0.03],  distance modulus [(m-M)$_{V}$ = 14.84],  metallicity [Fe/H] = -1.32 from \cite{Harris1996} (2010 edition) and with [Fe/H] = -1.39 $\pm$ 0.01 from \cite{2000AJ....119..840S}.

\section{Variables already known in NGC\,288}

Only 10 variable stars were known in NGC\,288. The first variable (V1) was discovered and identified by \cite{1943BAN.....9..397O} as a long-period semi-regular variable. V1 has a mean period of 103\,d. The second variable (V2) was identified by \cite{1977IBVS.1360....1H} as a RR Lyra with a period of 0.679\,d. 

\cite{1996A&AS..120...83K} found five additional short-period variable stars in the central part of the NGC\,288: one RR Lyra (V3) and four SX Phe stars (V4-V7). \cite{1997A&AS..125..337K} found three more new faint variables: two SX Phe stars (V8 and V9) and one contact binary (V10).  

\cite{2013AcA....63..429A} looked for new variables in NGC\,288, using nine observation nights during 2010 - 2013 at the 2.0-m telescope of the Indian Astronomical Observatory (IOA). They did not find additional variables and concluded that the census of RR Lyra stars was complete and if unknown SX Phe stars did exist, their amplitudes were smaller than the detection limit for their data.

In Table~\ref{tabvaria288} are shown the coordinates, types and magnitudes of the ten  known variables in NGC\,288.

\begin{table*}
\begin{minipage}{180mm}
\caption[]{ NGC\,288 known variable stars.} 
    \label{tabvaria288}
    \vspace{1em}
    \centering
\begin{tabular}{ l c c c c c c c } 
\hline\hline
Name & Type  & R.A.(J2000)    &  DEC.(J2000)  &     $\langle$U$\rangle$     &    $\langle$B$\rangle$      &    $\langle$V$\rangle$       &  $\langle$R$\rangle$   \\
  	 &	     &  (h:m:s)      &  (d:m:s)	    &   	   &	 	   &  			 & 	   \\
\hline 
V1$^{\ast}$ &  SM   &   00:52:41.13    &   -26:33:27.2      &   15.93  &	14.23  &	12.60  &	11.66   \\

V2$^{\ast\ast}$ &	RR Lyr	&	00:52:46.71    &   -26:34:07.0 	&	15.79  &	 15.48  & 15.66  &	15.28   \\
										
V3$^{\dagger}$ &	RR Lyr	&	00:52:40.28 	&  -26:32:28.4  	&	15.46  &	 15.59  &	15.20  &	15.15  \\
															
V4$^{\dagger}$ &	SX Phe	&	00:52:42.83 	&	-26:34:46.0		&	17.21  &	 17.56  &	17.31  &	17.16  \\
															
V5$^{\dagger}$ &	SX Phe	&	00:52:45.05 	&	-26:33:51.7		&	17.67  &	 17.83  &	17.55  &	17.60 	 \\
															
V6$^{\dagger}$	&	SX Phe	&	00:52:42.47 	&	-26:34:54.2	&	17.27  & 	17.62  &	 17.40  &	16.86  \\
															
V7$^{\dagger}$	&	SX Phe	&	00:52:41.46 	&	-26:33:59.3	&	18.02  &	18.28  &	17.95  &	17.80  \\
															
V8$^{\ddagger}$	&	SX Phe &	00:52:44.34 	&	-26:33:59.2	&	17.90  &	17.96  &	17.79  &	17.72 	 \\
															
V9$^{\ddagger}$ 	&	SX Phe	&   00:52:42.95 	&	-26:34:09.2	&	17.62  &	17.79  &	17.53  &	17.42  \\
															
V10$^{\ddagger}$ &  W-UMa binary & 	00:52:47.93     &   -26:33:01.4	&    19.55  &   19.81 	 &  19.21   &  18.78  \\
\hline
\end {tabular}
\\ $^{\ast}$\cite{1943BAN.....9..397O}, $^{\ast\ast}$\cite{1977IBVS.1360....1H}, 
 $^{\dagger}$\cite{1996A&AS..120...83K}, $^{\ddagger}$\cite{1997A&AS..125..337K}
\end{minipage}
\end{table*}

\section{Looking for new variables}

From our 300 time series images, all 11\,389 light curves were sorted using the \textit{E$_B$} index of variability in B filter, calculated with equation~\ref{indexE}. Based on the \textit{J} index of \cite{2007AJ....134..766K}, the \textit{E$_B$} index is defined as the average of the ​​of the deviations (in modulus) of the measured brightness in relation to the average brightness, 
normalized by the uncertainties in the measurements:

\begin{equation}
\label{indexE}
E_B = \frac{1}{N} \sum_{i=1}^N  \left| \frac{m_{Bi} - \overline{m}_B}{\sigma_{m_{Bi}}} \right|  \quad .
\end{equation}

\begin{figure}
\centering
\includegraphics[width=84mm,clip=]{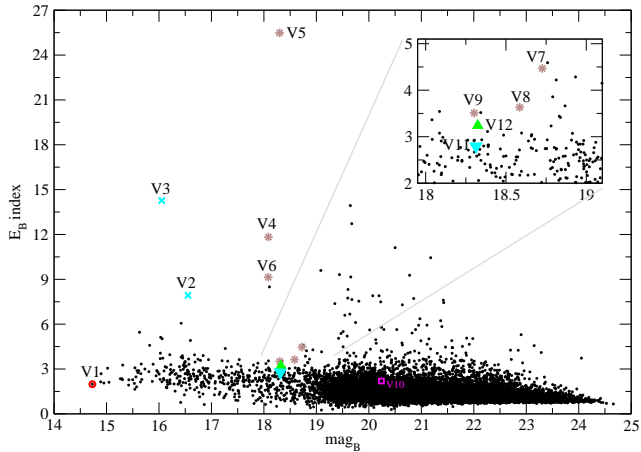}
\caption{ Distribution of index \textit{E$_B$} values with respect to the average B magnitude of NGC\,288 stars. Each different symbol represent a kind of variable: the variable V1 long period (circle), V2 and V3 are RR Lyr (crosses), V4 - V9 are SX Phe (asterisks) and the two new SX Phe variable stars V11 (triangle up) and V12 (triangle down).  }
\label{indexE288}
\end{figure}

In equation \ref{indexE}, $ m_{Bi} $ is the measured apparent magnitude in the B filter of the star in the \textit{i} image with uncertainty $\sigma_{m_{Bi}}$, and $\overline{m}_B$ is the average magnitude for the set of \textit{N} images. Statistically, for a light curve of a non-variable star with normal deviations in the brightness and well determined uncertainties, it is expected E$_B$ between $\sim$ 1.28 and $\sim$ 1.50. We considered all stars with E$_B > $ 1.4 as candidates to variables. This value corresponds to a probability of false alarm less than 5 per cent. The distribution of the E$_B$ indexes according to the B magnitude of the stars found in our images, is shown in Fig.~\ref{indexE288}. The positions of the known variables (V1 - V10) are indicated in the figure. As expected, for all the variables E$_B \gg $\,1.50.

Each light curve was visually inspected, and those that showed regular variability were selected. 
For these, we calculed the Fourier transform to look for pulsation frequencies with amplitudes higher than the 99 per cent confidence limit. When one or more frequencies were present, we removed the detected signal from the light curve (prewhitening), repeating all the process with the residual light curve in order to find other frequencies.

Subsequently, we modelled the variability in the light curve using the nonlinear multi-periodic function:
\begin{equation}
\label{multif}
I (t) = \sum_{k = 1}^n A_k \sin \left[2 \pi f_k (t - t_{max,k}) + \frac{\pi}{2}\right] \quad ,
\end{equation}
where, $f_k$ is the frequency, $A_k$ the amplitude and $t_{max,k}$ the time of maximum of the $k$th component. The non-linear fitting was done using the Levenberg-Marquardt method and the internal uncertainties were calculated from the full covariance matrix for the final fit.

\begin{figure*}
\begin{minipage}{115mm}
\centering
\includegraphics[width=115mm,clip=]{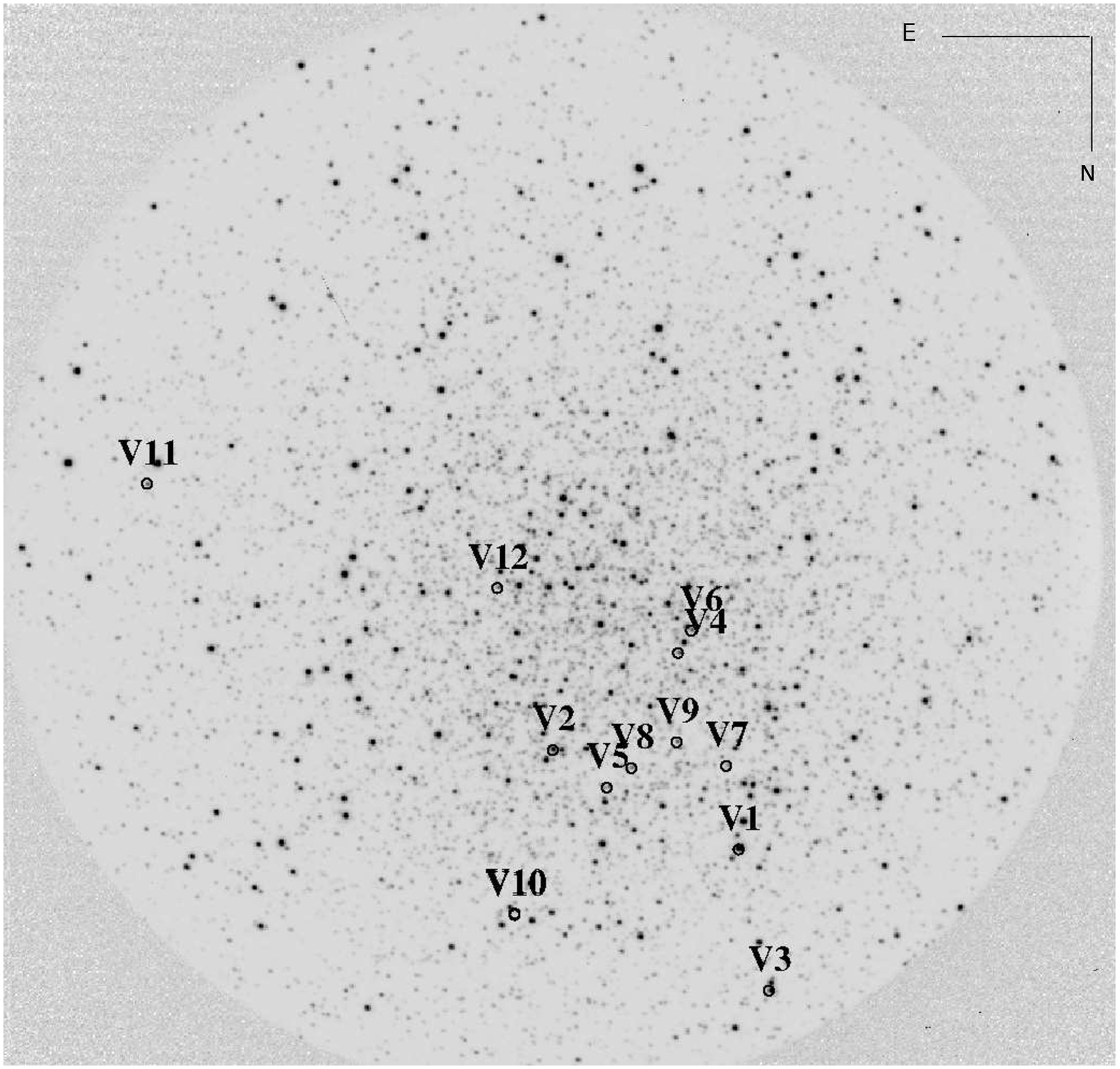}
\caption{ Position of the variable stars in the globular cluster NGC\,288. Image in B filter with 60\,s and diameter of 7.2\,arcmin centered at RA = $00^{h}52^{m}44^{s}.962$ and Dec.= $-26^{o}34\arcmin 57\arcsec.396$ (J2000). }
\label{pos288}
\end{minipage}
\end{figure*}

The results are shown in Tab.~\ref{tabvaria288b}. Position of variable stars in the globular cluster NGC\,288 are shown in Fig.~\ref{pos288}.

\begin{table*}
\begin{minipage}{180mm}
\caption[]{ Corrected version of Table~2 in \cite{2015MNRAS.447.2235M}. Detected frequencies  in the Fourier Transformes for the variables in NGC\,288.} 
    \label{tabvaria288b}
    \vspace{1em}
    \centering
\begin{tabular}{l l c c c c c c} \hline \hline 
Variable & Type 	      & E$_{\rm B}$ & Frequencies		           	& 		        	& Periods	            	& Amplitudes 	& T$_{{\rm max}_{{\rm i}}}$ 	\\
	 &		      &		& ($\mu$Hz)		   &		        &  (hours)	            	& (mma)		    		&  (sec)	\\ 
	 \hline 
	&		      &		&			   &		        &		                &		        	&		\\	
V3	& RR Lyr	      & 14.7	& $f_0$ = $\;$23.7$\pm$0.6 &		        & $P_0=$ 11.688$\pm$0.280	    	& 15.9$\pm$0.2			&  3\,479$\pm$505	\\
	&		      &		& $f_1$ = 174.1$\pm$3.9    &		        & $P_1=$1.596$\pm$0.037	    	& 1.0$\pm$0.2			&  5\,509$\pm$456 \\ 
\hline 
	&		      &		&			   &		        &		                &		       	 	&		\\ 
V4	& SX Phe	      & 11.8	& $f_0$ = 147.8$\pm$0.2    & 		        &$P_0=$	1.880$\pm$0.003	    	& 9.3$\pm$0.1			& 4\,492$\pm$18 	\\
	&		      &		& $f_1$ = 293.6$\pm$0.5	   & ($2\,f_0$)	&$P_1=$	0.946$\pm$0.002	    	& 3.8$\pm$0.1		   	& 2\,402$\pm$23		\\
	&		      &		& $f_2$ = 439.3$\pm$1.0	   & ($3\,f_0$)	&$P_2=$ 0.632$\pm$0.002	    	& 1.8$\pm$0.1		    	& 1\,708$\pm$34		\\
	&		      &		& $f_3$ = 585.0$\pm$2.5	   & ($4\,f_0$)	&$P_3=$	0.475$\pm$0.002	    	& 0.7$\pm$0.1	        	& 1\,314$\pm$62	\\
	&		      &		&			   &		        &		                &		        	&		\\
V5	& SX Phe	      & 25.8	& $f_0$ = 227.0$\pm$0.5    &		        &$P_0=$	1.224$\pm$0.003	    	&	12.8$\pm$0.2	    	&	2\,434$\pm$25	\\
	&		      &		& $f_1$ = 290.3$\pm$1.8    &		        &$P_1=$	0.957$\pm$0.006 	    	&	3.5$\pm$0.2 	    	&	2\,086$\pm$72	\\
	&		      &		&			   &		        &		                &		        	&		\\
V6	& SX Phe	      & $\;$9.1	& $f_0$ = 174.1$\pm$0.6	   &		        &$P_0=$ 1.595$\pm$0.005 		&	11.9$\pm$0.3	    	&	3\,642$\pm$38	\\
	&		      &		& $f_1$ = 345.8$\pm$1.1	   & ($2\,f_0$)	&$P_1=$ 0.803$\pm$0.003	        &	5.1$\pm$0.3 	    	&	1\,898$\pm$45	\\
&		      &		& $f_2$ = 513.9$\pm$2.7	   & ($3\,f_0$)	&$P_2=$ 0.540$\pm$0.003      	&	2.2$\pm$0.3	        &	1\,273$\pm$73	\\
	&		      &		&			   &		        &		                &		        	&		\\
V7	& SX Phe	      & $\;$4.5	& $f_0$ = 289.3$\pm$0.4    &		        &$P_0=$ 0.960$\pm$0.001		&  	2.4$\pm$0.1		&  2\,271$\pm$20		\\
	&		      &		&			   &		        &		                &		        	&		\\
V8	& SX Phe	      & $\;$3.6	& $f_0$ = 248.5$\pm$0.7    &		        &$P_0=$ 1.118$\pm$0.003		&	1.8$\pm$0.1 	    	&	2\,972$\pm$39	\\
	&		      &		&			   &		        &		                &		        	&		\\
V9	& SX Phe	      & $\;$3.5	& $f_0$ = 284.8$\pm$5.5    &		        &$P_0=$ 0.975$\pm$0.019	        &	0.9$\pm$0.4       	&	2\,715$\pm$191	\\	
	&		      &		& $f_1$ = 409.7$\pm$2.1    &	&$P_1=$ 0.678$\pm$0.004     	&	0.5$\pm$0.1      	&	$\;$2\,084$\pm$72	\\	
	&		      &		& $f_2$ = 299.2$\pm$3.4    &		        &$P_2=$ 0.9283$\pm$0.011           &	1.1$\pm$0.3	        &	$\;$979$\pm$126	\\
	&		      &		&			   &		        &		                &		        	&		\\		
\hline 
V11	& SX Phe	      & $\;$2.8	& $f_0$ = 273.4$\pm$1.7    &		        &$P_0=$	1.016$\pm$0.008    	&	1.7$\pm$0.1     	&	$\;$454$\pm$79	\\
	&		      &		& $f_1$ = 408.7$\pm$1.6    & 		&$P_1=$	0.680$\pm$0.003           &	1.2$\pm$0.1	        &	$\;$1\,271$\pm$57	\\	
	&		      &		& $f_2$ = 331.7$\pm$3.2    &		        &$P_2=$	0.838$\pm$0.009           &	0.6$\pm$0.1	        &	1\,712$\pm$137	\\	

	&		      &		&			   &		        &		                &		        	&		\\		
V12	& SX Phe	      & $\;$3.2	& $f_0$ = 266.0$\pm$3.3    &		        &$P_0=$	1.044$\pm$0.018       	&	0.7$\pm$0.1	        &	1\,959$\pm$213	\\
	&		      &		& $f_1$ = 400.8$\pm$2.1	   &	 	&$P_1=$	0.693$\pm$0.005       	&	0.4$\pm$0.1	    	&	$\;$730$\pm$84	\\	
	&		      &		&			   &		        &		                &		        	&		\\	\hline 
\end {tabular}
\end{minipage}
\end{table*}

For the V4 variable, we find three harmonics of the fundamental frequency, one for V5 and two for V6, as indicated in Tab.~\ref{tabvaria288b}.

\subsection{New variables}

\begin{figure*}
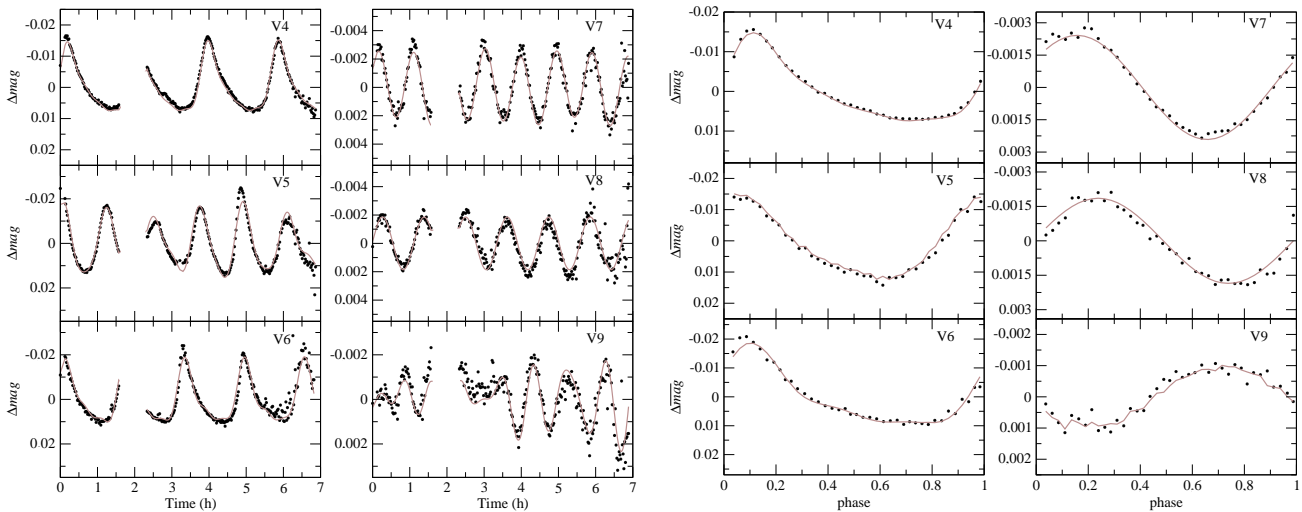
 
\centering
\begin{tabular}{cc}
\includegraphics[width=84mm,clip=]{SXPhe288.eps} & \includegraphics[width=84mm,clip=]{SXPhe288_phase.eps} \\
\end{tabular}
\caption{ Light curves (left) and folded light curves (right) of known variables in NGC\,288. To construct the folded light curves the main pulsation cycles were divided in 40 bins. The points represent the average of the measures inside each bin. }
\label{SXPhe288}
\end{figure*}

In the end, we retrieved all the six SX Phe (V4-V9), with amplitudes above the 99 per cent confidence level as shown in Table~\ref{tabvaria288}. Amplitudes are given in terms of milli-modulation in amplitude (1\,mma = 10$^{-3}$ma) and times of maximum in seconds, counted from the Julian date JD = 245\,6593.508576 (Barycentric Coordinate Time, BCT = 245 6593.513583). The light curves of the SX Phe (V4-V9) with the modelled variabilities by equation~\ref{multif} are shown in Fig.~\ref{SXPhe288}. 
Note that the models  fit the data very well.

\begin{figure*} 
\centering
\includegraphics[width=84mm,clip=]{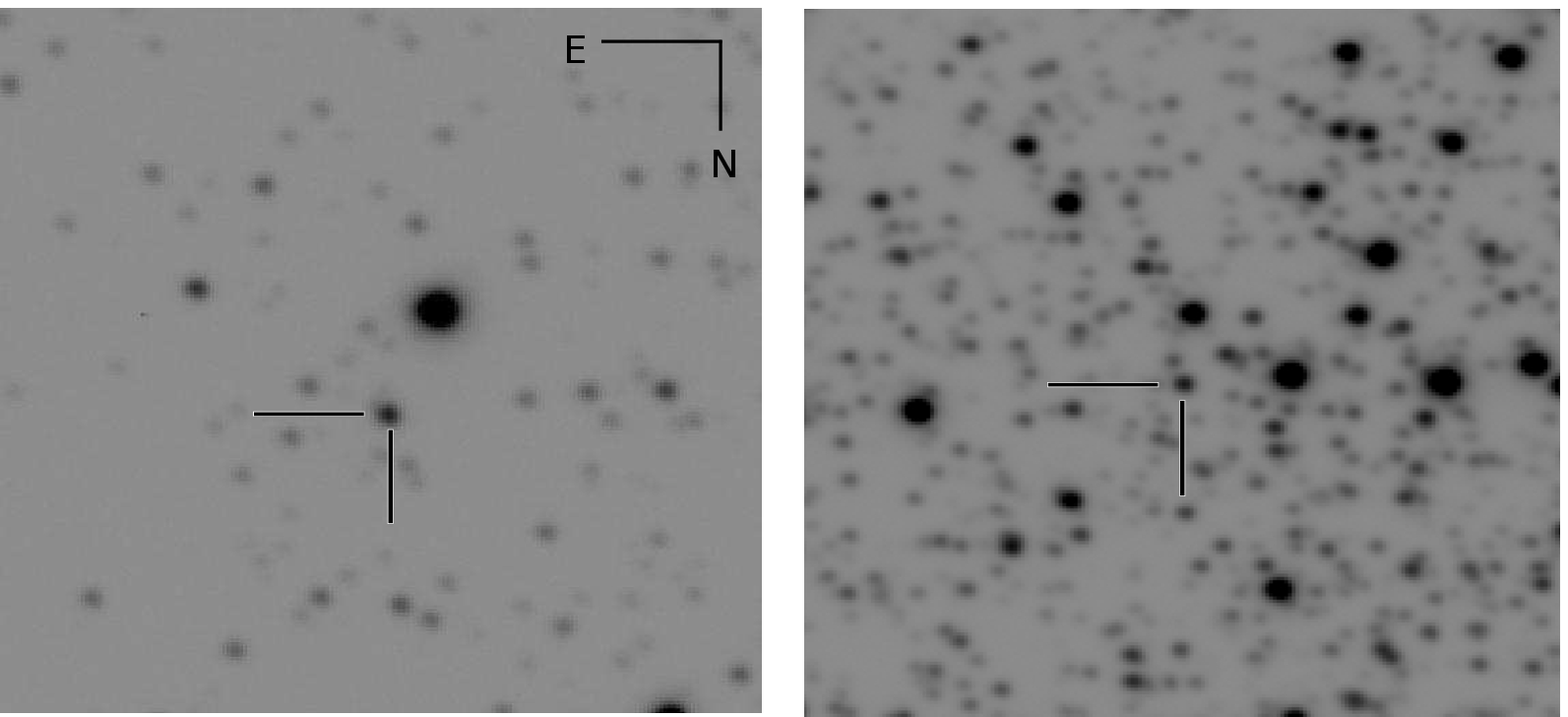}
\caption{ Finding charts for the NGC\,288 new variables: V11 (left) R.A. = 00:52:58.7, Dec. = -26:36:00.4 (J2000) and V12 (right) R.A. = 00:52:48.1, Dec. = -26:35:13.4. 
 Each chart is 30$\times$30\,arcsec. }
\label{desenho}
\end{figure*}

In addition, we found two new variables, named hereafter V11 and V12, with coordinates R.A.= 00:52:58.7, Dec.=-26:36:00.4 (J2000) and R.A.= 00:52:48.1, Dec.=-26:35:13.4 (J2000), respectively. Their positions are shown in the finding charts (Fig.~\ref{desenho}) and photometric parameters (Table~\ref{tabvaria}).

\begin{figure*}
\begin{minipage}{115mm}
\centering
\includegraphics[width=115mm,clip=]{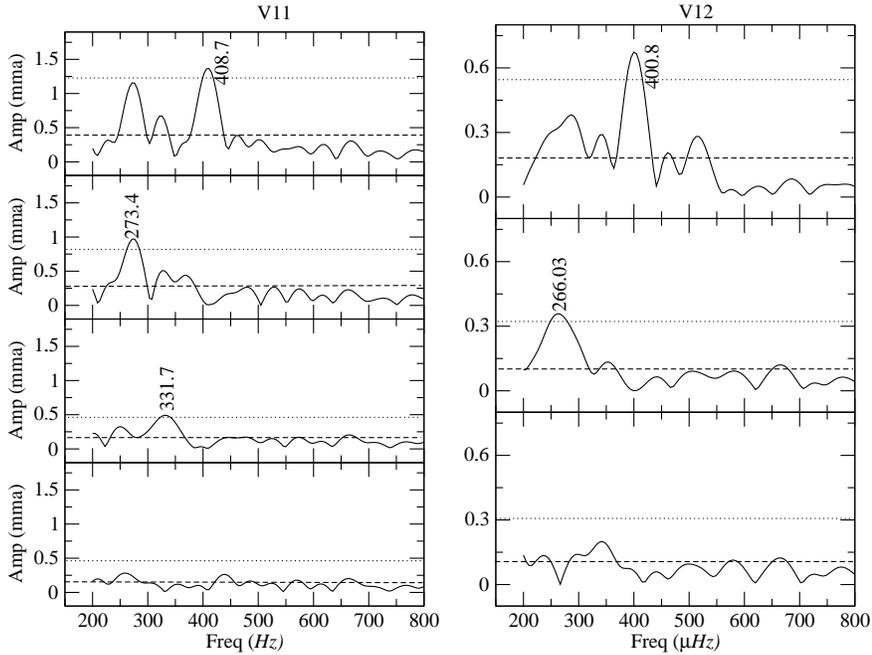}
\caption{ Prewhitening process in new SX Phe variable stars in NGC\,288: V11 (left) and V12 (right). The labels are frequencies in $\mu$Hz.  }
\label{dftNew}
\end{minipage}
\end{figure*}

\begin{figure*}
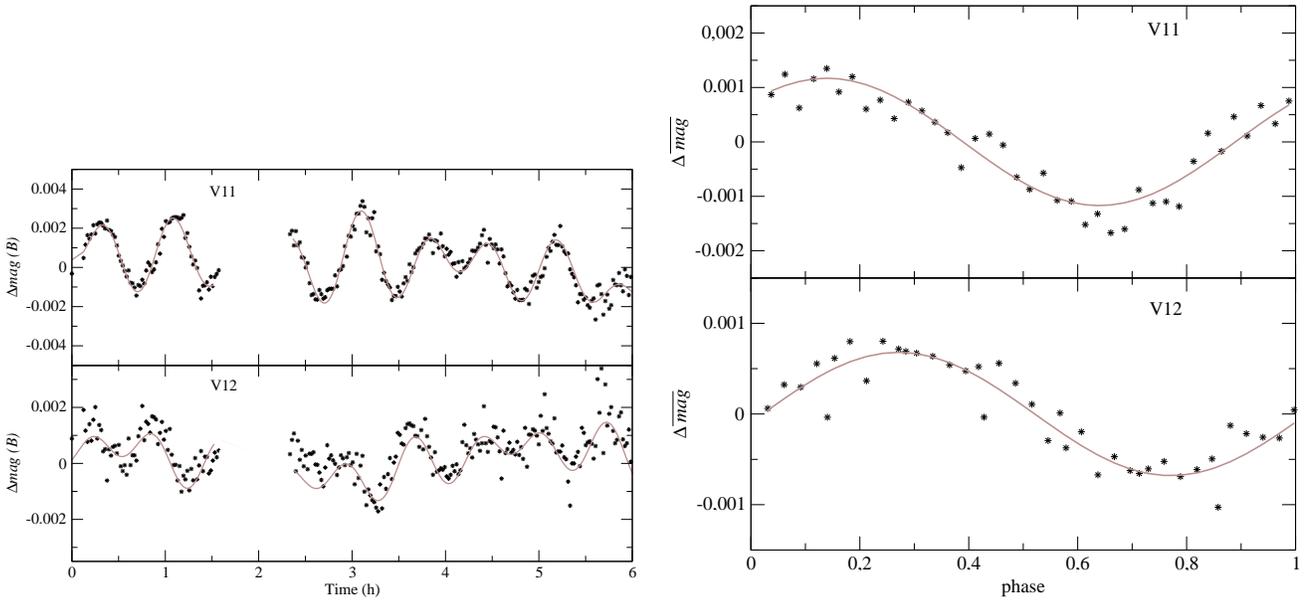
 
\centering
\begin{tabular}{cc}
\includegraphics[width=84mm,clip=]{newVar288syn.eps} & \includegraphics[width=84mm,clip=]{phase.eps} \\
\end{tabular}
\caption{ Light curves (left) and folded light curves (right) of the two new SX Phe variable stars in NGC\,288:  V11 (top) and V12 (bottom). The continuous curve is the fitted model described in the text. To construct the folded light curves the main pulsation cycles were divided in 40 bins. The points represent the average of the measures inside each bin.}
\label{newVar288syn}
\end{figure*}

For V11, we detected the frequencies: $f_0 $ = (273.4 $\pm$ 1.7)\,$\mu$\,Hz, $f_1$ = (408.7 $\pm$ 1.7)\,$\mu$\,Hz and $f_2$ = (331.7 $\pm$ 3.2)\,$\mu$\,Hz, corresponding to pulsation periods of $P_0 $ = (1.016 $\pm$ 0.008)\,h, $P_1$ = (0.680 $\pm$ 0.003)\,h and  $P_2$ = (0.838 $\pm$ 0.009)\,h, while, for V12 we found the frequencies, $f_0$ = (266.0 $\pm$ 3.3)\,$\mu$Hz and $f_1$ = (400.8 $\pm$ 2.1)\,$\mu$Hz, and pulsation periods of $P_0 $ = (1.044 $\pm$ 0.018)\,h and $P_1$ = (0.693 $\pm$ 0.005)\,h. 
The Fourier transforms for each step of the prewhitening process are shown in Fig.~\ref{dftNew} and the light curves modelled by equation~\ref{multif} and the folded light curves are shown in Fig.~\ref{newVar288syn}. 

The pulsation periods of V11 and V12 are in the range of the SX Phe variables and both are in the SX Phe region in the CMDs, as shown in Fig.~\ref{CMDMart}. 
As V7, V8 and V9, the new SX Phe present low modulation in amplitude (less than 1.2\,mma) and, therefore, low $E_B$ index, what place the all five stars in the same region in the $E_B$ diagram (Fig.~\ref{indexE288}).

\section{SX Phoenicis stars}

SX Phoenicis (SX Phe) variable stars have been discovered in galaxies usually belonging to their globular clusters. Similar to RR Lyra stars and Cepheids, the SX Phe are useful distance indicators. 

Light curves of SX Phe variables in globular clusters are used to search for double-mode oscillation. Some SX Phe stars pulsate only in the fundamental mode while in other cases both the fundamental and first-overtone modes as observed by \cite{1997PASP..109.1221M}. 
There is a predicted period ratio between the fundamental mode and the first overtone ($P_1/P_0$), and between the secondary period oscillation and first overtone ($P_2/P_1$) for known SX Phe stars in globular clusters \citep{2000ASPC..210..391P}. The ratios are $P_1/P_0$ = 0.773 and $P_2/P_1$ = 0.803 for solar-metallicity models \citep{2011AJ....142..110M}. \cite{1998ApJ...507..818G} suggests that the behavior of the pulsation properties of SX Phe are partially affected by metal content. This is supported by \cite{2001ApJ...554.1124S} who found differences in the pulsation period when comparing metal-poor models to metal-rich ones.

SX Phe stars have masses between $\sim$ 1.0 and 1.3 M$_\odot$ \citep{2011AJ....142..110M} belonging to the metal-poor group of pulsating variables in the lower instability strip among the $\delta$ Scuti stars \citep{1991A&AS...91..347R}, but while $\delta$ Sct stars are Population I stars, SX Phoenicis variables are of Population II. They are in the CMD blue-straggler region, with short periods between 0.8132\,h and 3.021\,h (corresponding to frequencies approximately between 90 and 340\,$\mu$Hz), the majority with periods of 1.90\,h \citep{2011AJ....142..110M}.

The SX Phe-type variables are not yet fully understood by stellar evolution theory. 
The hypothesis that most readily explains the origin of SX Phe stars is that they are mergers from pre-existing close binaries \citep{2000JRASC..94..124B}. 
It is assumed that they arose by the merger of two main-sequence stars, in a close binary. Additional astrophysics theories dealing with the production of the variables are necessary \citep{2011AJ....142..110M}.

\begin{table}
\caption[]{Photometric parameters of the two new variable, V11 and V12.} 
    \label{tabvaria}
    \vspace{1em}
    \centering
	\begin{tabular}{ c  c  c  } 
	\hline\hline
			&	V11		&	V12	 	\\
			&				&				\\
	\hline	
R.A.(h:m:s)			& 	00:52:58.7	&	00:52:48.1	\\
Dec. (g:m:s)			&	-26:36:00.4	&	-26:35:13.4	\\
				&				&				\\
$\Delta$ B 	&	0.03		&	0.025		\\
				&				&				\\
$\langle$U$\rangle$			&	17.65$\pm$0.01			&		17.67$\pm$0.01		\\
$\langle$B$\rangle$		    &	17.81$\pm$0.02			&	    17.78$\pm$0.02			\\
$\langle$V$\rangle$			&	17.55$\pm$0.02			&		17.56$\pm$0.01			\\
$\langle$R$\rangle$		    &	17.41$\pm$0.02			&		17.36$\pm$0.02		\\
				&				&				\\		
(U-B)			&	-0.16$\pm$0.02		&		-0.11$\pm$0.02	\\	
(B-V)			&	0.26$\pm$0.03		&		0.22$\pm$0.02		\\	
(V-R)			&	0.14$\pm$0.02		&		0.20$\pm$0.02	\\	
\hline
    \end {tabular}
\end{table}

SX Phe stars are known to present a Period-Luminosity (P-L) relation, which can be used as distance indicator for globular clusters \citep{2003AJ....125.3165J}. 

\begin{figure}
\centering
\includegraphics[width=84mm,clip=]{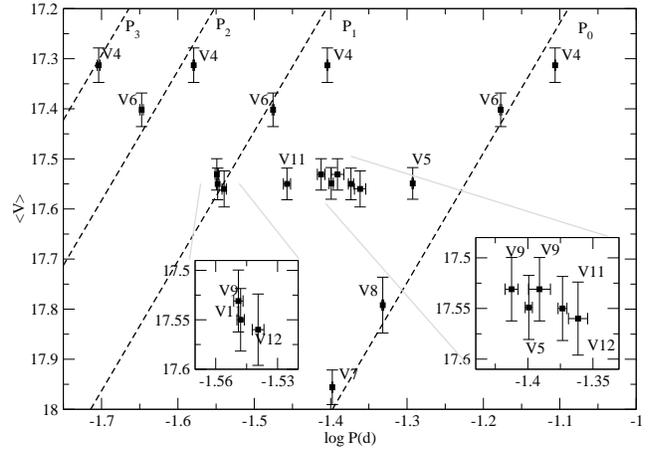}
\caption{ Period-brightness relations (dashed lines) for NGC\,288 SX Phe stars for the fundamental mode ($P_0$) and the three first harmonics $P_1$, $P_2$ and $P_3$.}
\label{MAG_PER288}
\end{figure}

In Fig.~\ref{MAG_PER288} the period-brightness relations are indicated by the dashed lines.
The brightness is given in terms of the average magnitude in the V band and the periods in days. 
We use \cite{2013AcA....63..429A} who adopted the SX Phe PL relation derived by \cite{2011MNRAS.416.2265A} in M\,53.% ($M_V =2.916 log P(d)-0.898 $).
Using the periods of V4-V8 identified in Table~\ref{tabvaria288b} as fundamental periods and the five identified as harmonics in V4 and V7 [using $P_k / (k+1)$] we obtain the period-brightness relation:

\begin{equation}
\label{fmode}
V =  (-2.59 \pm 0.18) \log P_0 + (14.38 \pm 0.23) \quad .
\end{equation}

\noindent 
The period-brightness relations for the first three overtones are directly derived from the above equation,
replacing $P_0$ with $P_k / (k+1)$:

\begin{equation}
  V = (-2.59 \pm 0.18) \log P_1 + (13.56 \pm 0.24) \quad ,
 \label{eq_pbP1}
 \end{equation}
 
 \begin{equation}
 V = (-2.59 \pm 0.18) \log P_2 + (13.18 \pm 0.25)  
 \end{equation}
 
 \noindent
 and
 
 \begin{equation}
 V = (-2.59 \pm 0.18) \log P_3 + (12.89 \pm 0.25) \quad .
 \end{equation}

\noindent 

It is interesting to note what happens with V5, V9, V11 and V12, all them with V magnitude very close to 17.55. V9, V11 and V12 present $P_1$ consistent with the period-brightness relation for first overtone, but $P_0$ displaced ($3\sigma$ or more) in relation to the predicted fundamental periods. 
For V5, this happens with both periods. 
A possible explanation is locking by resonance between pulsation modes.
The double mode of V5 and V9 was observed and discussed already by \cite{2013AcA....63..429A}. For V9 as well as for V11 and V12, we did not detect the oscillation fundamental mode in our light curves.

\noindent 

Using distance modulus (m-M)$_o$ = 14.57 $\pm$ 0.08 (see Sec. 3), we obtain the period-luminosity relations:

\begin{equation}
\label{fmodeB}
M_V = (-2.59 \pm 0.18) \, \log P_0 + (- 0.19 \pm 0.23)
\end{equation}

\noindent
and 

\begin{equation}
M_V = (-2.59 \pm 0.18) \, \log P_1 + (0.50 \pm 0.25) \quad .
\end{equation}

\section{Summary and discussions}

Our main objective in this work was to search for new variable stars in the globular cluster NGC\,288. 
Each B-filter light curve was visually inspected according the \textit{E$_B$} index and for those which showed significant variability we calculated the Fourier transform in order to find the pulsations frequencies. The frequencies were used as input data to fit a nonlinear multiperiod function to refine the fit of frequencies and calculate the amplitudes and phases. 
This fit allowed us to also improve the determination of periods and frequencies of the already known variable stars. 

Furthermore, we found two new variables stars, named here, V11 and V12. We classify the new variable stars as SX Phoenicis due to their characteristic periods and amplitude of pulsation of this type of stars. They are also located in the SX Phe region in CMD. The location of the two new variables in the CMD also shows that they belong of the globular cluster NGC\,288.

The two new variables are in the same region of the E$_B$ diagram as the already know SX Phe V7, V8 and V9. This is because the five variables have small oscillation amplitudes and the same order of magnitude and therefore must share other similar physical properties.

We also obtained the main parameters of the globular cluster NGC\,288 comparing the mean ridge line from the observed CMDs using the \textit{Dartmouth Stellar Evolution Database} grid of isochrones and we made a correction for the CMD with the V magnitude and (B-V) color using the differential-reddening map. As result, the mean metallicity is [Fe/H] = -1.3 $\pm$ 0.1, the age model is 13.5\,Gyr, the mean reddening is [E(B-V)] = 0.02 and the distance modulus for the best agreement is 14.72 $\pm$ 0.01 for the V-filter, with a distance of 8.8 $\pm$ 0.1\,kpc. 

We use the period-luminosity relation for SX Phe stars to determine the distance of the globular cluster using all periods obtained with our settings for eight variable stars (known and new discoveries). The two P-L relations were determined, one for the fundamental mode and the other for the first-overtone. 
The average distance we determine by the P-L relation is 8.71 $\pm$ 0.20\,kpc. Through a weighted average of all the distances (derived from the P-L relations and isochrones results) we determine the best distance for the globular cluster NGC 288 as 8.8 $\pm$ 0.3\,kpc, which is similar to that of \cite{2013AcA....63..429A} of 8.9 $\pm$ 0.3\,kpc, also using a SX Phe P-L relation.

NGC\,288 is a interesting case where detected SX Phe stars outnumber RR Lyrae.
As RR Lyrae, SX Phe are good for determining distances. With this work, we discovered two new variables, increasing the number of SX Phe known in globular clusters.

\section*{Acknowledgments}

We thank the referee for important comments and suggestions.
We thank the SOAR support team for help with the data acquisition.
We acknowledge financial support from the Brazilian Institution CAPES, CNPq and FAPERGS/PRONEX.

\end{document}